\documentclass[11pt]{article}
\usepackage{amsfonts,amssymb, amsmath}
\usepackage[english]{babel}

\def\nn{\nonumber }
\def\bq{ \begin{equation} }
\def\eq{ \end{equation} }
\def\ben{ \begin{eqnarray} }
\def\en{ \end{eqnarray} }

\newtheorem{prop}{Proposition}

\newtheorem{defi}{Definition}

\newtheorem{re}{Remark}
\newenvironment{rem}{\begin{re} \rm }{\end{re}}

\begin{document}


\title{On bi-integrable natural
Hamiltonian systems on Riemannian  manifolds }
\author{ A V Tsiganov \\
\it\small
St.Petersburg State University, St.Petersburg, Russia\\
\it\small e--mail:  andrey.tsiganov@gmail.com}

\date{}
\maketitle

\begin{abstract}
 We introduce the concept of natural Poisson bivectors, which  generalizes the Benenti approach to construction of  natural  integrable systems on   Riemannian manifolds and  allows us to  consider almost the whole  known  zoo of integrable systems in framework of bi-hamiltonian geometry.
\end{abstract}

\vskip0.1truecm

\section{Introduction}
\setcounter{equation}{0}

Let us consider a natural integrable by Liouville system on a Riemannian  manifold $Q$ for which the  Hamilton function
\bq \label{nat-h}
H=T+V=\sum_{i,j=1}^n \mathrm g_{ij}\,p_i\,p_j+V(q_1,\ldots, q_n)\,
\eq
is the sum of the geodesic Hamiltonian $T$ and potential energy $V$. Integrability means that there are functionally independent integrals of motion $H_1=H,H_2,\ldots,H_n$ in  involution
\[\{H_i,H_j\}=\langle dH_i,PdH_j\rangle=0\,,\]
with respect to the canonical Poisson brackets defined by the following Poisson bivector
\bq\label{can-p}
P=\left(
  \begin{array}{cc}
    0 & \mathrm{I} \\
    -\mathrm{I} & 0
      \end{array}
\right)\,.
\eq
Among   integrable systems we want to pick out a family of   bi-integrable systems for which:
\begin{itemize}
  \item There is a second Poisson bivector $P'$ compatible with  $P$, i.e.
\bq\label{m-eq1}
[P,P']=0,\qquad [P',P']=0,\qquad
\eq
where $[.,.]$ is the  Schouten bracket.
  \item Integrals of motion $H_1,\ldots,H_n$ are in  bi-involution
      with respect to  both Poisson brackets
      \bq\label{m-eq2}
      \{H_i,H_j\}=\{H_i,H_j\}'=0.
      \eq
\end{itemize}

First  example of bi-integrable systems are bi-hamiltonian systems.  The concept of bi-ha\-mil\-to\-ni\-an vector fields was introduced firstly by Magri   studying the Korteweg-de-Vries equation in order to explain integrability
of soliton equations from the standpoint of classical analytical mechanics \cite{mag78}.   However,  for the overwhelming  majority of  known natural integrable  systems on  Riemannian  manifolds the  Hamiltonians $H$ (\ref{nat-h})  give rise to  non bi-hamiltonian vector fields $X=PdH$.  The natural  obstacle  for existence of the bi-hamiltonian vector fields in finite-dimensional case is discussed in  \cite{br}.

The second special but more  fundamental example of bi-integrable systems are separable systems, for which there  exist $n$ separation relations of the form
\begin{equation}
\label{seprel}
\phi_i(u_i,v_i,H_1,\dots,H_n)=0\ ,\quad i=1,\dots,n\ ,
\qquad\mbox{with }\det\left[\frac{\partial \phi_i}{\partial H_j}\right]
\not=0\> .
\end{equation}
Here  $u=(u_1,\ldots,u_n)$ and $v=(v_1,\ldots,v_n)$ are canonical variables of separation, $\{u_i,v_j\}=\delta_{ij}$.
The proof of the fact that any separable system  is  a bi-integrable system with respect to second Poisson bivector
\[
P'=\left(  \begin{array}{cc}
    0 & \mathrm{F} \\
    -\mathrm{F} & 0
      \end{array}
\right)\,,\qquad \mathrm{F}=\mbox{diag}\Bigl(f_1(u_1,v_1),\ldots, f_n(u_n,v_n)\Bigr)
\]
labeled by $n$ arbitrary functions $f_k$ on variables of separation may be found in \cite{ts07}. The main problem is how to describe these bivectors in terms of initial physical variables or, equivalently, how to determine variables of separation for a given natural Hamilton function on $Q$.

This problem is solved for the quadratic in momenta integrals of motion  $H_1,\ldots,H_n$  separable in  orthogonal coordinate systems on $Q$, see \cite{ben92,ben93,ben97,bm03,bl09,cr03,fp02,hms05,imm00,st} and references within. In this case initial physical  coordinates are related with coordinates of separation by the point transformation
\bq\label{point-trans}
q_i=g_i(u_1,\ldots,u_n),\qquad p_i=\sum_{j=1}^n h_{ij}(u_1,\ldots,u_n)v_j\,.
\eq
The first  algebraic condition for separability  systems  with quadratic in momenta  integrals of motion has been found by St\"{a}ckel  \cite{st}.  Then in  \cite{lc04} Levi-Civita  proved that a Hamilton-Jacobi equation $H(q, p) = E$ admits a separated solution if and only if the  separability conditions or separability equations of Levi-Civita  are identically satisfied
  \[
\partial_i H\partial_j H\partial^i\partial^jH+\partial^i\partial^jH\partial_i\partial_jH-\partial_iH\partial^jH\partial^i\partial_jH-
\partial^iH\partial_jH\partial_i\partial^jH=0\,,\qquad i,j=1..n\,,
  \]
here $\partial_k=\partial/\partial u_k$ and $\partial^k=\partial/\partial v_k$\,.  Using transformations (\ref{point-trans}) it is easy to rewrite  Levi-Civita equations as polynomial equations of fourth degree in the momenta $p_1,\ldots,p_n$ and to note that  fourth-degree homogeneous part of the Levi-Civita equations depends only on the geodesic Hamiltonian  $T$. Such as Levi-Civita equations  must be identically satisfied for all admissible values of $p_1,\ldots,p_n$  it means that:
\begin{itemize}
  \item the separation of the geodesic equation is a necessary condition for the separation at $V\neq 0$;
  \item the study of the geodesic separation
plays a prominent role.
\end{itemize}
As was shown by Eisenhart (for an orthogonal case at $\mathrm g_{ij} = 0$ for $i\neq j$) and by Kalnins and Miller for generic case, the geodesic separation is related to the existence of Killing vectors and Killing tensors of order two \cite{eis34, km80,km81}.
These ideas are nicely embraced by the geometric notion of  Killing webs discussed in \cite{ben92,ben93,ben97}.

In framework of the geometric  Benenti theory  it is possible to construct a basis of  Killing tensors by  coordinate independent algebraic procedure starting with  special  tensor $L$ with the following properties:
 \begin{enumerate}
   \item  $L$ is a conformal Killing tensor of gradient type,
   \item the Nijenhuis torsion of $L$ vanishes,
   \item $L$ has pointwise simple eigenvalues.
 \end{enumerate}
According to  \cite{bl09,fp02,imm00,tur92}, these conditions entail that we can define the second Poisson bivector
 \bq\label{p2-ben}
 P'=  \left(
      \begin{array}{cc}
        0 & L_{ij} \\
        \\
         -L_{ij}\qquad&\displaystyle \sum_{k=1}^n\left(\dfrac{\partial L_{ki}}{\partial q_j}-\dfrac{\partial L_{kj}}{\partial q_i}\right)p_k
      \end{array}
    \right)\,,
 \eq
and that  eigenvalues of the recursion operator $N=P'P^{-1}$ are the desired variables of separation.  Some algorithms and software for calculation of the Benenti tensor $L$ starting with a given natural Hamilton function on the Riemannian  manifold $Q$ of constant curvature  may be found in \cite{ts05,hms05,rw03}.

In this note we consider natural Hamiltonians (\ref{nat-h}) on  $\mathbb R^{2n}$ with unit metric tensor
\bq\label{nham-r}
H=\sum_{i=1}^n p_i^2+V(q_1,\ldots q_n)\,,\qquad \mathrm g_{ij}=\delta_{ij}\,,
\eq
and the corresponding \textit{natural} Poisson bivectors
  \bq\label{p2-gen}
 P'=
 \left(
      \begin{array}{cc}
        \displaystyle \sum_{k=1}^n\left(\dfrac{\partial \Pi_{jk}}{\partial p_i}-\dfrac{\partial \Pi_{ik}}{\partial p_j}\right)q_k  & \Pi_{ij} \\
        \\
         -\Pi_{ji}\quad&0
      \end{array}
    \right)
+
 \left(
      \begin{array}{cc}
        0 & \Lambda_{ij} \\
        \\
         -\Lambda_{ji}\quad &\displaystyle \sum_{k=1}^n\left(\dfrac{\partial \Lambda_{ki}}{\partial q_j}-\dfrac{\partial \Lambda_{kj}}{\partial q_i}\right)p_k
      \end{array}
    \right)\,,
 \eq
which are \textit{separable} on geodesic and potential parts. Here geodesic Hamiltonian $T$ and geodesic matrix $\Pi$ depend only on  momenta $p_1,\ldots,p_n$ and $\Pi$ has zero  Nijenhuis torsion as a tensor field on $\mathbb R^n$ with these coordinates.  Potential $V$ and potential matrix $\Lambda$ depend  only on coordinates $q_1,\ldots, q_n$ and the  Nijenhuis torsion of $\Lambda$ on  configurational space $Q=\mathbb R^n$ is equal to zero.

This paper  belongs mostly to so-called "experimental" mathematical physics.
Our main aim is to present some of the most striking examples of matrices $\Pi,\Lambda$ and to discuss the yet unsolved problems:
\begin{itemize}
  \item invariant definition and construction of $\Pi$ and $\Lambda$ without an unpretentious solution of the equations (\ref{m-eq1}) in the special coordinate system fixed by unit metric tensor (\ref{nham-r});
  \item finding the  integrals of motion in  bi-involution from $P'$;
  \item calculation of  separation variables;
  \item definition of similar bivectors on another Riemannian  manifolds.
\end{itemize}
Note that at $\Pi=0$ we have exactly the same problems, which have been partially solved in framework of the Benenti theory. In fact the Benenti recursion procedure concerns with a  special subclass of the St\"ackel systems, which contains for instance the classical Jacobi separation of the geodesic flow on an asymmetric ellipsoid   \cite{bm03,ben05}. Another solutions of these problems are related with  lifting  of the St\"ackel systems to bi-Hamiltonian systems of Gel'fand-Zakharevich type using extension of the initial phase space \cite{bl09,fp02,imm00}.
However, all these solutions are relatively simple because  momenta $p_j$ are linear functions on   $v_j$ (\ref{point-trans}) and, therefore, our  Hamilton function $H$ has   natural form simultaneously in physical variables $(p,q)$ and in variables of separation  $(u,v)$.

In generic case at $\Pi\neq 0$ for the present we do not have any satisfactory solutions of these problems. Some efforts of applying  the Killing theory  to integrable systems with higher order integrals of motion have been made in \cite{mst03}. Particular solutions of the other problems by  brute force method may be found in  \cite{mpt,ts06,ts07b,ts08,ts10,ts10a,ts09v}.

The paper is organized as follows. In Section 2 the concept of natural Poisson bivectors  on the Riemannian manifolds is briefly reviewed.  The Toda lattices and  rational Calogero-Moser systems  illustrate  possible applications of this  concept.  In Section 3  the problem of classification  bi-integrable systems  on low-dimensional Euclidean spaces is treated.
The Henon-Heiles system, the system with quartic potential, the Holt-like system and some new bi-integrable systems are discussed as well. In Section 4  we introduce natural Poisson bivectors on the sphere $\mathbb S^n$. At $n=2$  we show natural bivectors  associated with  the Kowalevski top, Chaplygin system and Goryachev-Chaplygin top.
The possible generalizations of natural Poisson bivectors  are discussed in  final Section.

\section{Natural Poisson bivectors on the Riemannian manifolds}
\setcounter{equation}{0}
Let $Q$ be an $n$-dimensional  Riemannian manifold.  Its  cotangent bundle $T^*Q $ is naturally
  endowed with canonical invertible Poisson bivector $P$, which  has the standard form (\ref{can-p}) in fibered coordinates ($p,q$) on  $T^*Q$.

\begin{defi}
A Poisson bivector $P'$ on $T^*Q$ has a natural form if it is a sum of
the geodesic Poisson bivector $P'_T$ and the  potential Poisson bivector
defining by torsionless (1,1) tensor field $\Lambda(q_1,\ldots, q_n)$ on $Q$
\bq\label{n-p}
P'= P'_T+\left(
      \begin{array}{cc}
        0 & \Lambda_{ij} \\
        \\
         -\Lambda_{ji}\quad &\displaystyle \sum_{k=1}^n\left(\dfrac{\partial \Lambda_{ki}}{\partial q_j}-\dfrac{\partial \Lambda_{kj}}{\partial q_i}\right)p_k
      \end{array}
    \right)\,,
\eq

 \end{defi}
 In fact, here we assume that  bi-integrability of the geodesic  motion is a necessary condition for  bi-integrability in generic case at $V\neq 0$  and, therefore,  natural bivector $P'$  has to remain the Poisson bivector at  $\Lambda=0$. It allows us to classify  all  possible geodesic solutions $P'_T$ of the  equations (\ref{m-eq1}) and then to add to them various consistent potential parts.

The main  result of the our experiments is that all the known  geodesic Poisson bivectors $P'_T$  have the following form
\bq\label{n-p2}
P'_T=\left(
      \begin{array}{cc}
       \partial_p \Pi  & \Pi \\
        \\
         -\Pi^\top\quad& \partial_q \Pi
      \end{array}
    \right)\,.
\eq
Here entries of geodesic matrix $\Pi$ are the  homogeneous second order polynomials in momenta
\bq\label{pi2-sph}
\Pi_{ij}=\sum_{k,m=1}^n c_{ij}^{km}(q) p_kp_m\,,
\eq
similar to the geodesic Hamiltonian $T$ (\ref{nat-h}) , whereas antisymmetric tensors $\partial_{p} \Pi$ and $\partial_{q} \Pi$ are given by (\ref{p2-gen}) or by more complicated expressions   considered in  Section \ref{toda}, Section \ref{s-sphere} and Section \ref{hh-sep}.

\begin{prop}
Natural Poisson bivectors $P'$ (\ref{n-p},\ref{n-p2}) are unambiguously determined by  a pair of    geodesic  and potential $n\times n$ matrices $(\Pi,\Lambda)$  on $2n$-dimensional space $T^*Q$.
\end{prop}
Our definition of the  natural Poisson bivectors $P'$ drastically depends on a choice  of coordinate system.  We hope that  further inquiry of invariant  geometric relations between  metric tensor $G$ and matrices  $\Pi,\Lambda$  allows us to get more invariant and rigorous  mathematical  description of these objects.

There is  a family of \textit{integrable} potentials $V$, which may be added to a given geodesic Hamiltonian $T$ in order to get integrable Hamiltonian $H=T+V$.  By analogy, we  define a family of \textit{compatible} potential matrices $\Lambda$,
which may be added  to a given geodesic matrix $\Pi$ in order to get   natural Poisson  bivector $P'$  compatible with  $P$.
\begin{defi}
Potential matrix  $\Lambda$ is compatible with geodesic matrix $\Pi$  if the
natural Poisson bivector $P'$ (\ref{n-p},\ref{n-p2})   satisfies the equations (\ref{m-eq1}),  so that   $P - \lambda P'$  is   a Poisson bivector for each $\lambda$.
\end{defi}

\begin{rem}
In some sense  we have to waive previous the principle that  geodesic motion is completely independent from  potential. In fact  for the same geodesic Hamiltonian $T$ we have many different geodesic matrices $\Pi$ compatible with various potential matrices $\Lambda$, i.e. our choice of $\Pi$ depends on potential $V$.
\end{rem}

\subsection{ Integrals of motion versus variables of separation}
Below we  suppose that natural Poisson bivector  $P'$ (\ref{n-p}) is always  compatible with canonical bivector $P$, so that   the phase space  $T^*Q$ becomes bi-hamiltonian manifold endowed   with the hereditary recursion operator
\[N=P'P^{-1}\,.\]
 In general,  there are three different occasions:
\begin{enumerate}
  \item recursion operator produces the necessary number of  integrals of motion;
  \item recursion operator generates variables of separation instead of integrals of motion;
    \item recursion  operator produces only  some part of the integrals of motion or variables of separation.
\end{enumerate}

In the first case, the  traces of powers of the recursion operator $N$ are functionally independent constants of motion
\bq\label{ham-gen}
H_1=T+V=\dfrac{1}{2}\,\mbox{trace}\, N\,,\qquad H_k=\dfrac{1}{2k}\mbox{trace}\, N^k\,, \quad k=2,\ldots,n,
\eq
and our natural  Hamiltonian $H$ (\ref{nat-h}) is directly determined by $\Pi$ and $\Lambda$:
\[H=H_1\,,\qquad  T=\sum_{i=1}^{n} \Pi_{ii}\,,\qquad\mbox{and}\qquad V=\sum_{i=1}^{n}  \Lambda_{ii}\,.\]
This Hamiltonian defines natural bi-Hamiltonian system on $T^*Q$.

In the  second  case, the  traces of powers of the recursion operator $N$ remain  functionally independent  constants of motion  for an \textit{auxiliary}   bi-Hamiltonian system on $T^*Q$, which differs for our target system with  Hamiltonian  $H$ (\ref{nat-h})
\[H\neq \dfrac{1}{2}\,\mbox{trace}\, N\,,\qquad  T\neq  \sum_{i=1}^{n} \Pi_{ii}\,,\qquad\mbox{and}\qquad V\neq \sum_{i=1}^{n}  \Lambda_{ii}\,.\]
In this situation we will treat  eigenvalues $u_j$ of   $N$
\bq\label{dn-var}
B(\lambda)=\Bigl(\,\det(N-\lambda\mathrm I)\,\Bigr)^{1/2}=(\lambda-u_1)(\lambda-u_2)\cdots
(\lambda-u_n)\,,
\eq
 as  separation variables for a huge family of \textit{separable} bi-integrable systems on $T^*Q$ associated with various separated relations  (\ref{seprel}). Of course, this construction will be justified only if we are capable to obtain  desired Hamilton functions $H$  (\ref{nat-h}) from  (\ref{seprel}).

In  third case recursion operators $N$
is degenerate and produces only  a part of  integrals of motion or variables of separation. We can leave this difficulty  and  get the necessary number of  functionally independent integrals of motion  using some additional assumptions about the so-called control matrices $ F$   defined by
\bq\label{f-mat}
P'dH_i=P\sum_{j=1}^n   F_{ij}\,dH_j,\qquad i=1,\ldots,n.
\eq
According to \cite{ts06}, we can fix some special forms of  $F$ and try to get the corresponding natural Hamiltonians $H_1$ and natural Poisson bivectors $P'$ simultaneously.

Now we are going to  illustrate the first and the third opportunities by  examples  of the $n$-body Toda lattice and of the rational Calogero-Moser system, respectively.  Construction of  separation variables are considered in  Section \ref{sep1} and Section \ref{sep-gen}.

\subsection{The Toda lattices}\label{toda}
Let us start with a well-known second Poisson tensor for the open Toda lattice associated with $\mathcal A_n$ root system \cite{das89}:
\bq\label{toda-gen} \widehat{P}=\sum_{i=1}^{n-1}
e^{q_i-q_{i+1}}\dfrac{\partial}{\partial
p_{i+1}}\wedge\dfrac{\partial}{\partial p_{i}} +\sum_{i=1}^n
p_i\dfrac{\partial}{\partial q_{i}}\wedge\dfrac{\partial}{\partial
p_{i}}+\sum_{i<j}^n \dfrac{\partial}{\partial
q_{j}}\wedge\dfrac{\partial}{\partial q_{i}}.
\eq
This bivector is different from the natural Poisson bivector defined by (\ref{n-p}-\ref{n-p2}). Nevertheless,  using recursion operator $\widehat{N}=\widehat{P}P^{-1}$, it is easy to get a quadratic in momenta  Poisson bivector
\bq\label{p-toda}
P'=\widehat{N}\widehat{P}\,,
\eq
 which has the natural form of (\ref{n-p}-\ref{n-p2}) if we put
\bq\label{pi-toda}
\Pi=\mbox{diag}(p_1^2,\ldots,p_n^2)\,,\qquad \partial_q\Pi=0\,,\qquad
(\partial_p\Pi)_{ij}=
\left\{
  \begin{array}{cl}
    \dfrac{1}{2}\left(\dfrac{\partial \Pi_{ii}}{\partial p_i}+\dfrac{\partial \Pi_{jj}}{\partial p_j}\right), &i<j; \\
    0, \quad& i=j;\\
    -\dfrac{1}{2}\left(\dfrac{\partial \Pi_{ii}}{\partial p_i}+\dfrac{\partial \Pi_{jj}}{\partial p_j}\right), & i>j;
  \end{array}
\right.
\eq
and if $n\times n$ potential tensor $\Lambda=-E\,A$ is a product of two antisymmetric matrices with entries
\bq\label{l-atoda}
 E_{ij}=\left\{
          \begin{array}{rl}
            1,\quad & i<j; \\
            0,\quad & i=j; \\
           -1,\quad &i>j;
          \end{array}
        \right.
 \qquad\mbox{and}\qquad A_{i,i+1}=a\,e^{q_i-q_{i+1}}\,.
\eq
In this case the recursion operator $N=P'P^{-1}$ produces the necessary number of  integrals of motion $H_k$ defined by  (\ref{ham-gen}) and the Hamilton function has the natural form
\[H_1=\dfrac{1}{2}\,\mbox{trace}\,N=\sum_{i=1}^n p_i^2+2a\sum_{i=1}^{n-1} e^{q_i-q_{i+1}}\,.\]
 In order to get variables of separation we have to introduce another linear in momenta Poisson bivector \cite{ts07}, which may be rewritten in  natural form as well.

\begin{rem} In fact matrix $A$ (\ref {l-atoda}) is the well-known second  matrix  in the Lax equation $\dot  L=[ L,A]$ for the open Toda lattice.  In framework of the group theoretical settings of integrable systems the Lax matrices are viewed as a coadjoint orbits of a Lie algebras. We believe that natural bivector $P'$ (\ref{p-toda})  has  transparent algebro-geometric justification, similar to compatible bivectors from \cite{bb02}.
\end{rem}

The Poisson bivectors  for the  Toda lattices associated with $\mathcal{BC}_n$ and $\mathcal D_n$ root systems have the natural form of  (\ref{n-p}-\ref{n-p2}) if
\[
\Pi=\mbox{diag}(p_1^2,\ldots,p_n^2)\,,\quad \partial_q\Pi=0\,,\qquad
(\partial_p\Pi)_{ij}=\left\{
                       \begin{array}{cl}
                         \dfrac{\partial \Pi_{ii}}{\partial p_i},\quad& i<j; \\
                         0,\quad & i=j; \\
                        -\dfrac{\partial \Pi_{ii}}{\partial p_i},\quad &i>j;
                       \end{array}
                     \right.
\]
and if $n\times n$ potential parts are given by
\bq
\begin{array}{ll}
\mathcal{BC}_n\qquad &  \Lambda=-(\mathrm I+2\widetilde{E})A+be^{q_n}\,B+ce^{2q_n}\,C \\ \\
\mathcal D_n\qquad &  \Lambda=-(\mathrm I+2\widetilde{E})A+de^{q_{n-1}+q_{n}}\,D,\qquad b,c,d,\in\mathbb R\,.
\end{array}
\eq
where  $\mathrm I$ is the unit matrix,  $A$ is given by (\ref{l-atoda}) and   $\widetilde{E}$ is a strictly upper triangular matrix
\[
\widetilde{E}_{ij}=\left\{\begin{array}{cc}
                           1 & i<j \\
                           0 & i\geq j
                         \end{array}\right.\,.
\]
Matrices $B,C$ and $D$ have non-zero entries only in the last columns:
\[
B=\left(
    \begin{array}{cccc}
      0 &  \cdots & 0&1 \\
       \vdots &  \ddots & \vdots& \vdots \\
      0 &  \cdots &0& 1 \\
      0 &  \cdots & 0&1 \\
    \end{array}
  \right)\,,\qquad C=\left(
    \begin{array}{cccc}
      0 &  \cdots & 0&2\\
       \vdots &  \ddots &\vdots&  \vdots \\
      0 &  \cdots &0& 2 \\
      0 &  \cdots &0& 1 \\
    \end{array}
  \right)\,,\qquad
  D=\left(
    \begin{array}{ccccc}
      0 &  \cdots & 0&2&2 \\
      \vdots &  \ddots & \vdots& \vdots& \vdots \\
      0 &  \cdots &0& 2& 2 \\
      0 &  \cdots &0& 2& 1 \\
      0 &  \cdots & 0&-1&0 \\
    \end{array}
  \right)\,.
\]
As above,  recursion operators $N$ generate  integrals of motion $H_k$ (\ref{ham-gen}) and  Hamilton functions have the natural form
\[
\begin{array}{ll}
\mathcal{BC}_n\qquad &\displaystyle H_1=\sum_{i=1}^n p_i^2+2a\sum_{i=1}^{n-1} e^{q_i-q_{i+1}}+be^{q_n}+ce^{2q_n}\\ \\
\mathcal D_n\qquad &\displaystyle  H_1=\sum_{i=1}^n p_i^2+2a\sum_{i=1}^{n-1} e^{q_i-q_{i+1}}+2de^{q_{n-1}+q_n}\,.
\end{array}
\]
These natural  Poisson bivectors in $(p,q)$-variables  have been obtained in \cite{dam04},   whereas bi-hamiltonian structures for the periodic Toda lattices and  construction of the separation  variables  are discussed in \cite{ts07a,ts08t}.

\begin{rem}
The  "relativistic" modification of the  natural Poisson bivector (\ref{n-p}) associated with
 relativistic $n$-body Toda model  \cite{rag89}  is  considered  in Section (\ref{r-toda}).
\end{rem}

\subsection{The Calogero-Moser system}\label{cal}
The bi-Hamiltonian formulation  of the Calogero-Moser system can be  found in  \cite{ped10, nut01,MagMar}. We present new and very simple natural Poisson bivector  (\ref{p2-gen}), which is different form these known Poisson brackets  expressed directly in terms of integrals $H_k$ .

The $n$-particle rational Calogero-Moser model  associated with the root system $\mathcal A_n$ is defined by the Hamilton function
\begin{equation}
H=\dfrac{1}{2}\sum_{i=1}^{n}p_{i}^{\;2}-{a^2}\sum_{i\neq
j}^n \frac{1}{(q_{i}-q_{j})^{2}}  \label{ham-cal}
\end{equation}
where $a$ is a coupling constant.  The second  natural Poisson bivector $P'$  (\ref{p2-gen}) for this system   is defined by
 symmetric  geodesic matrix
\bq\label{geo-cal}
\Pi= p \otimes p\,,\qquad \Pi_{ij}=p_ip_j\eq
and potential matrix $\Lambda$ with entries
\bq\label{pot-cal}
\Lambda_{ij}=q_i\sum_{k\neq j}^n \dfrac{a^2}{(q_j-q_k)^3}\,.
\eq
 In this case recursion operator $N=P'P^{-1}$ generates only  the Hamilton function
 \bq \label{cal-deg}
 \mbox{trace}\,N^k=2(2H)^{k}\,\qquad\mbox{and}\qquad  P\,dH=P'd\ln H\,.
 \eq
and we have so-called irregular bi-Hamiltonian manifold. Nevertheless, integrals of motion $H_k=\dfrac{1}{k!}\,\mbox{trace}\,\mathcal L^k$ obtained from the standard Lax matrix
 \[
\mathcal L
=\left(\begin{matrix} p_1 & \frac{a}{q_{1}-q_{2}} & \cdots & \frac{a}{q_{1}-q_{n}}\\
\frac{a}{q_{2}-q_{1}} & p_2 & \ddots & \vdots\\
\vdots & \ddots & \ddots & \vdots\\
\frac{a}{q_{n}-q_{1}} & \frac{a}{q_{n}-q_{2}} & \cdots & p_n
        \end{matrix}\right)\ .
\]
 are in  bi-involution (\ref{m-eq2})  with respect to the Poisson brackets defined by $P$ (\ref{can-p})
 and $P'$  (\ref{geo-cal}-\ref{pot-cal}).  Besides these $n$  integrals of motion the rational
Calogero-Moser system admits $n-1$ additional functionally
independent integrals of motion $K_{m}$
\[
K_{m}= m g_{1}H_{m}- g_{m} H_{1},  \qquad
g_{m} = \frac{1}{2} \, \left\{\sum_{i=1}^{n} q_{j}^{\;2}, H_{m} \right\}\,,\qquad
m=2,...,n\,.
\]
All  these  integrals of motion $H_k$ and $K_m$ may be obtained from
the Hamilton function $H=H_2$  (\ref{ham-cal}) as polynomial solutions of the  following  equations
\bq\label{cal-int}
P\,dH=\dfrac{1}{k}\,P'\,d\ln H_k=\dfrac{1}{m-1}\,P'\,d\ln K_m\,.
\eq
In  Section \ref{cm-example} we discuss
solutions of  the equations (\ref{cal-int})  associated with another natural Poisson bivectors and
other bi-integrable systems.

\begin{rem}
 We suppose that  trigonometric (elliptic) Calogero-Moser systems  and their generalizations associated with  other root systems may be associated with natural Poison bivectors, see \cite{ts06} at n=2 . In order to describe the Ruijsenaars-Schneider model we can try to  introduce another Poisson bivector $P'$  similar to relativistic Toda case, see Section \ref{r-toda}.
\end{rem}

\section{Natural bivectors   on low-dimensional Euclidean spaces.}
\setcounter{equation}{0}
Now let us come back to the Poisson manifold $ R^{2n}$  endowed with the natural Poisson bivector (\ref{p2-gen}). In this case the corresponding   Poisson bracket $\{.,.\}'$ looks like
\ben
&&\{q_i,p_j\}'=\Pi_{ij}+\Lambda_{ij}\,,\quad
\{q_i,q_j\}'=\sum_{k=1}^n\left(\dfrac{\partial \Pi_{jk}}{\partial p_i}-\dfrac{\partial \Pi_{ik}}{\partial p_j}\right)q_k\,,\nn\\
&&\{p_i,p_j\}'=\sum_{k=1}^n\left(\dfrac{\partial \Lambda_{ki}}{\partial q_j}-\dfrac{\partial \Lambda_{kj}}{\partial q_i}\right)p_k\,.
\en
 The geodesic Hamiltonian $T$   is  second order  homogeneous polynomial in momenta, so  we are going to  suppose that  entries of $\Pi$ are   second order homogeneous  polynomials  in momenta as well.  This assumption allows as to  get a lot of  natural  Poisson bivectors $P'$ (\ref{p2-gen}) compatible with canonical bivector $P$ and describe the corresponding bi-integrable systems.  For brevity  we will not consider the complete classification and  restrict ourselves by discussing only more interesting examples  for $n=2,3$. Some other examples may be found in \cite{mpt,ts06}.

\subsection{Integrals of motion via recursion operator}
In this section we suppose that  geodesic Hamiltonian $T$ and potential $V$ are directly determined by $\Pi$ and $\Lambda$  (\ref{ham-gen}).

\vskip0.2truecm
\par\noindent
\textbf{Case 1}: Let us start with the  non-degenerate geodesic matrix
\bq\label{pi-1}
 \Pi^{(1)}=\dfrac{1}{2}\left(
      \begin{array}{cc}
        p_1^2+\frac{1}{2}p_2^2 & 0 \\
        \frac{1}{2}p_1p_2 & \frac{1}{2}p_2^2
      \end{array}
    \right)\,,\qquad\mbox{so that}\qquad T=\dfrac{p_1^2+p_2^2}{2}\,.
\eq
There are  some potential matrices $\Lambda$ compatible with it, for instance:
{\setlength\arraycolsep{2pt}
\ben
\Lambda^{(1)}&=&\left(
          \begin{array}{cc}
            \left(\dfrac{3c_1q_2}{8}+\dfrac{c_2}{8}\right)q_1^2+c_1q_2^3+c_2q_2^2+c_3q_2\qquad &
             \dfrac{c_1q_1^3}{16}+\left(\dfrac{3c_1q_2^2}{2}+c_2q_2+\dfrac{c_3}{2}\right)q_1
             \\ \\
            -\dfrac{c_1q_1^3 }{32}& c_1q_2^3+c_2q_2^2+c_3q_2
          \end{array}
        \right)\,,\nn\\
        \nn\\
        \nn\\
V^{(1)}&=&\dfrac{c_1}{8}\,q_2(3q_1^2+16q_2^2)+c_2\,\left(2q_2^2
+\dfrac{q_1^2}{8}\right)+2c_3q_2\,,\label{case-1}
\en
and
\ben
\Lambda^{(1')}&=&\left(
          \begin{array}{cc}
           \dfrac{c_1q_1^4}{4}+\left(3c_1q_2^2+c_2\right)\dfrac{q_1^2}{2}
           +c_1q_2^4+c_2q_2^2+\dfrac{c_3}{q_2^2}  &\dfrac{c_1q_1^3q_2}{2}+\left(2c_1q_2^3+c_2q_2-\dfrac{c_3}{q_2^3}\right)q_1

             \\ \\
             -\dfrac{c_1q_1^3q_2}{4}& c_1q_2^4+c_2q_2^2+\dfrac{c_3}{q_2^2}
          \end{array}
        \right)\,,\nn\\
        \nn\\
        \nn\\
V^{(1')}&=&\dfrac{c_1}{4}(q_1^4+6q_1^2q_2^2+8q_2^4)
+\dfrac{c_2}{2}(q_1^2+4q_2^2)+\dfrac{2c_3}{q_2^2}\,.\label{case-1p}
\en}
The second integrals of motion $H_2=\mbox{trace}\,N^2$ are  fourth order polynomials in momenta:
\bq\label{case-1-h2}
H_2^{(1)}=p_1^4+\dfrac{q_1^2(3c_1 q_2+c_2)}{2}p_1^2-\dfrac{c_1q_1^3}{2}p_1p_2
-\dfrac{q_1^4}{32}\Bigl(c_1^2(6q_2^2+q_1^2)+c_1(8c_3+4c_2q_2)-2c_2^2\Bigr)
\eq
and
\ben\label{case-1p-h2}
H_2^{(1')}&=&\left(p_1^2+c_2q_1^2\right)^2
+c_1q_1^2\Bigl((q_1^2+6q_2^2)p_1^2+q_1^2p_2^2-4q_1q_2p_1p_2\Bigr)
+\frac{4c_1c_3q_1^4}{q_2^2}\\
&+&
\frac{1}{4}c_1q_1^4(q_1^2+2q_2^2)(c_1q_1^2+2c_1q_2^2+4c_2)\,.
\nn
\en
These integrable systems  were found by using the weak-Painlev\'{e} property of equation of motion  and the direct search of  fourth order polynomial integrals of motion, see  \cite{gdr84,h87}.
\begin{rem}
The Henon-Heiles system with potential $V^{(1)}$ and the system with fourth order potential $V^{(1')}$ admit various integrable generalizations \cite{gdr84,h87}, which are considered in the Section \ref{hh-gen}.
\end{rem}
\textbf{Case 1g}:
At $n=3$ the immediate generalization of $\Pi^{(1)}$ (\ref{pi-1}) looks like
\bq\label{pi-13}
\Pi^{(1g)}=\dfrac12 \left(
   \begin{array}{ccc}
    \dfrac{p_1^2}{3}+\dfrac{p_2^2}{2} & 0  & \dfrac{\sqrt{2}p_1p_2}{3}+\dfrac{2p_1p_3 }{3}\\
    \\
    \dfrac{p_1p_2}{2} & \dfrac{p_2^2}{2} & \dfrac{\sqrt{2}\,p_1^2}{6} \\
    \\
    \dfrac{\sqrt{2}p_1p_2}{2} & -\dfrac{\sqrt{2}p_1^2}{3} & \dfrac{2p_1^2}{3}+p_3^2
   \end{array}
 \right)\,,\qquad T=\dfrac{p_1^2+p_2^2+p_3^2}{2}\,.
\eq
One of the potential matrices $\Lambda$ compatible with $\Pi^{(1g)}$  is equal to
\ben
\Lambda^{(1g)}&=&c_1
\left(
  \begin{array}{ccc}
    \frac{q_2(9q_1^2+8q_2^2)}{2} & \frac{q_1(q_1^2+8q_2^2)}{4} & \frac{\sqrt{2}q_1(3q_1^2+16q_2^2-4q_3^2+8\sqrt{2}q_2q_3)}{8} \\
    \\
     -\frac{9q_1^3}{8}& 4q_2^3  & -\frac{3\sqrt{2}q_2^2(q_2+\sqrt{2}q_3)}{4} \\
     \\
     \frac{3\sqrt{2}q_1(3q_1^2+4q_2^2+2q_3^2+2\sqrt{2}q_2q_3)}{4}&
     \frac{3\sqrt{2}q_1^2(4q_2+\sqrt{2}q_3)}{4} &  3q_1^2q_2+3\sqrt{2}q_1^2q_3+\sqrt{2}q_3^3
  \end{array}
\right)\nn
\nn\\
\nn\\
&+&c_2\left(
  \begin{array}{ccc}
    3q_1^2+8q_2^2 & 8q_1q_2 & 4\sqrt{2}q_1q_2 \\ \\
    0 & 8q_2^2 &-\frac{3\sqrt{2}}{2}q_1^2 \\
     6(\sqrt{2}q_2+q_3)q_1 & 3\sqrt{2}q_1^2 & 4q_3^2+6q_1^2
  \end{array}
\right)
+c_3\left(
  \begin{array}{ccc}
    q_2 & \frac{q_1}{2} &  \frac{\sqrt{2}q_1}{4} \\
        0 & q_2 & 0 \\
        \frac{3\sqrt{2}q_1}{4} & 0 & \frac{\sqrt{2}q_3}{2}
  \end{array}
\right)\,,
\nn
\en
so that
\ben
V^{(1g)}=c_1\left(8q_2^3+\dfrac{15}{2}q_1^2q_2+3\sqrt{2}q_3q_1^2+\sqrt{2}q_3^3\right)+
c_2(16q_2^2+9q_1^2+4q_3^2)+c_3\left(2q_2+\dfrac{\sqrt{2}}{2}q_3\right)\,.\nn
\en
In this case the  integrals of motion $H_2=\mbox{trace}\,N^2$ and $H_3=\det N$ are the fourth and sixth order polynomials in momenta, such that few pages are necessary to write them down.

\begin{rem}
At $c_{2,3}=0$ this potential is equivalent to potential $V_{10}$ in \cite{p09a}.  Similar Poisson bivectors may be constructed for another potentials from \cite{p09a}  and for $n$ dimensional generalizations of the Henon-Heiles systems and systems with quartic potentials \cite{dor86}.
\end{rem}

\subsection{Integrals of motion via control matrices}\label{cm-example}
In \cite{ts06} we have obtained some natural Poisson bivectors using the  Lenard and Fr\"{o}benius  control matrices.
Now we  present some new examples of two-dimensional bi-integrable systems   associated with  degenerate  control matrix
\bq\label{f-2}
F=\left(
    \begin{array}{cc}
      H_1 & 0\\
      \varkappa^{-1}H_2 & 0 \\
    \end{array}
  \right)\,,\qquad \varkappa \in\mathbb R\,,
\eq
It means that Hamiltonian $H_1$ is  the solution of  the equation
\[
X=P_1dH_1= P'\,d\,\ln H_1\,,
\]
whereas $H_2$  is  the solution of  the  equation depending on rational parameter  $\varkappa$
\bq\label{def-h2d}
X=P_1dH_1=\varkappa\, P'\,d\,\ln H_2\,,
\eq
similar to the  equations for the  Calogero-Moser system (\ref{cal-int})\,.

An algebraic construction of such two-dimensional Hamilton functions $H_1$, additional integrals of motion $H_2$ and  natural  Poisson bivectors $P'$ (\ref{p2-gen})  has been proposed in \cite{mpt}.
\vskip0.2truecm
\par\noindent
\textbf{Case 2}: Let us consider the degenerate symmetric  matrix associated with the Calogero-Moser systems (\ref{geo-cal})
\bq\label{pi-2}
 \Pi^{(2)}=\dfrac{1}{2}\left(
      \begin{array}{cc}
        p_1^2 & p_1p_2 \\
        p_1p_2 & p_2^2
      \end{array}
    \right)\,,\qquad\mbox{such that}\qquad T=\dfrac{p_1^2+p_2^2}{2}\,.
\eq
There are  some other  potential matrices $\Lambda$ compatible with it, for instance,
\ben\label{case-2}
\Lambda^{(2)}&=&\dfrac{c_1}{(q_1^2+q_2^2)^2}\left(
          \begin{array}{cc}
            q_1^2 & q_1q_2 \\
            \\
            q_1q_2 & q_2^2
          \end{array}
        \right)\,,
\qquad
V^{(2)}=\dfrac{c_1}{q_1^2+q_2^2}=\dfrac{c_1}{r^2}\,\\
\nn\\
 H_1^{(2)}&=&T+V^{(2)}\,,\qquad H_2^{(2)}=2(p_1q_2-p_2q_1)^2\,H_1^{(2)}\,,\nn
\en
and
\ben\label{case-2p}
\Lambda^{(2')}&=&\left(
          \begin{array}{cc}
             \dfrac{c_1}{q_1^2}+\dfrac{(d+2)c_2q_2^d}{2q_1^{d+2}}& -\dfrac{dc_2  q_2^{d-1}}{2q_1^{d-3}} \\
             \\
           \dfrac{c_1q_2}{q_1^3}+\dfrac{(d+2)c_2q_2^{d+1}}{q_1^{d+3}} &-\dfrac{dc_2 q_2^d}{q_1^{d+2}}
          \end{array}
        \right)\,,
\qquad
 V^{(2')}=\dfrac{c_1}{q_1^2}+\dfrac{c_2\,q_2^d}{q_1^{d+2}}\,,\\
\nn\\
 H_1^{(2')}&=&T+V^{(2)}\,,\qquad
H_2^{(2')}=\left((p_1q_2-p_2q_1)^2+\dfrac{2c_1\,q_2^2}{q_1^2}
+(-1)^{d+1}\dfrac{2c_2\,q_2^d(q_1^2+q_2^2)}{q_1^{d+2}}\right)\,H_1^{(2')}\,.\nn
\en
In  both cases second integrals of motion $H_2$ were found as the
solutions of the equation (\ref{def-h2d}) at $\varkappa=1$. Let us note that the matrix  $\Lambda^{(2')}$ (\ref{case-2p}) is a particular case of matrices  fixed by
\[
\Lambda_{12}=\dfrac{1}{q_1^2}\,\Phi\left(\frac{q_2}{q_1}\right)\,,\]
where $\Phi$ is an arbitrary function  \cite{ts06}.
\vskip0.2truecm
\par\noindent
\textbf{Case 2g}:  System  (\ref{pi-2} -\ref{case-2}) has an obvious $n$-dimensional counterpart
\bq\label{pi-23}
\Pi^{(2g)}_{ij}=\dfrac{p_ip_j}2 \,,\qquad
\Lambda^{(2g)}_{ij}=\dfrac{c_1q_iq_j}{(\sum q_i^2)^2}
\,,\qquad H_1=\dfrac{1}{2}\sum_{i=1}^n p_i^2+\dfrac{c_1}{\sum q_i^2}\,.
\eq
\begin{rem}
According to  \cite{ts09}, at $n=2$ there are a lot of integrable deformations of the centrally
symmetric potential $V=\dfrac{c_1}{r^2}$, for which  the forth and the sixth order polynomial integrals $H_2$ do  not the products of the Hamilton function  $H$ and  the second order polynomials as in (\ref{case-2}). It will be interesting to study similar deformations at $n>2$.
\end{rem}

\par\noindent
\textbf{Case 3}: Now let us consider another metric and  degenerate  non-symmetric matrix
\bq\label{pi-3}
 \Pi^{(3)}=\dfrac{1}{2}\left(
      \begin{array}{cc}
       a p_1p_2 & bp_2^2 \\
        ap_1^2 &  bp_1p_2
      \end{array}
    \right)\,,\qquad\mbox{such that}\qquad T=(a+b){p_1p_2}\,.
\eq
It is easy to prove that matrix $\Lambda$ may be added to $\Pi^{(3)}$ if and only if
\[
\Lambda_{12}=q_1^{-\frac{2b}{a}}\Phi(q_2q_1^{-\frac{b}{a}})\,.
\]
For instance, if $\Phi(z)=z^d$ and $\gamma=(2b+a+bd)$ then
\ben
\Lambda^{(3)}&=&\left(
  \begin{array}{cc}
  \dfrac{c_1\gamma}{a+b}\,q_2^{d+1}\,q_1^{-\dfrac{\gamma}{a}}+c_2q_1^{-\frac{a+b}{a}}
        & -\dfrac{ac_1(d+1)}{a+b}\,q_2^d\,q_1^{-\frac{b(d+2)}{a}}\\
     \\
     \\
   \dfrac{c_1\gamma b}{a(a+b)}\,q_2^{d+2}\,q_1^{-\frac{\gamma}{a}-1}
   +\dfrac{c_2b}{a}\,q_2\,q_1^{-\frac{a+b}{a}-1} & -\dfrac{bc_1(d+1)}{a+b}\,q_2^{d+1}\,q_1^{-\frac{\gamma}{a}}
  \end{array}
\right)\,,\nn\\
\nn\\
\nn\\
V^{(3)}&=&c_1\,q_1^{-\gamma/a}\,q_2^{d+1}+c_2\,q_1^{-(a+b)/a}\,.\label{case-3}
\en
In order to get an additional integral of motion $H_2$ we have to use the equation (\ref{def-h2d}) because the  recursion operator $N$ is degenerate.  Depending  on values of $a,b$ and $d$ second integral of motion  $H_2$  may be second, forth or sixth order polynomial in momenta. For instance,
we present some examples with forth order polynomial integrals of motion
\ben
&V=q_1^3q_2^{-\frac{9}{5}}\,,\qquad
&H_2= 4p_1^4-10\left(3p_1^2q_1^2-30p_1p_2q_1q_2+25p_2^2q_2^2\right)q_2^{-4/5}+225q_1^4q_2^{-8/5}\,,\nn\\
\nn\\
&V=q_1^2q_2^5\,,\qquad
&H_2=16p_1^3(p_1q_1-p_2q_2)+4p_1q_1q_2^6(p_1q_1-2p_2q_2)+q_2^8\left(p_2^2-q_1^3q_2^4\right)\,,\nn\\
\nn\\
&V=q_1^2q_2^{-\frac{7}{4}}\,,\qquad
&H_2=2p_1^3(p_1q_1-p_2q_2)-q_2^{-\frac{3}{4}}\left(13p_1^2q_1^2-80p_1p_2q_1q_2+64p_2^2q_2^2\right)+64q_1^3q_2^{\frac{-3}{2}}\,,\nn
\en
and six order polynomial integrals of motion
\ben
&V=q_1^{-\frac{2}{3}}q_2^{-\frac{5}{6}}\,,\quad
&H_2=2q_1^{-\frac{2}{3}}\left(p_1q_1-p_2q_2\right)^2\left(
q_2^{\frac{1}{6}}-2p_1q_1^{\frac{2}{3}}(p_1q_1-p_2q_2) \right)+q_1^{-\frac{1}{3}}q_2^{-\frac{1}{3}}\,,\nn\\
\nn\\
&V=q_1^{\frac{3}{2}}q_2^{-\frac{7}{2}}\,,\quad
&H_2=q_1p_1^6-p_2q_2p_1^5-\dfrac{5q_1^{3/2}p_1^4}{2q_2^{5/2}}
+\dfrac{3q_1^2p_1^2}{2q_2^5}
+\dfrac{3q_1p_1p_2}{4q_2^4}+\dfrac{p_2^2}{4q_2^3}-\dfrac{q_1^{5/2}}{8q_2^{15/2}}\,.\nn
\en
Other examples may be found in \cite{mpt}.  It will be interesting to find  a generic expression for all second integrals of motion $H_2$ associated with the potential matrix (\ref{case-3}).
\vskip0.2truecm
\par\noindent
\textbf{Case 4}: Let us consider a non-degenerate  geodesic matrix
\bq\label{pi-4}
 \Pi^{(4)}=\dfrac{1}{2}\left(
      \begin{array}{cc}
       a p_1p_2 & -\dfrac{a-b}{2}\,p_2^2 \\
       \\
        \dfrac{a-b}{2}\,p_1^2 &  bp_1p_2
      \end{array}
    \right)\,,\qquad\mbox{such that}\qquad T=(a+b){p_1p_2}\,.
\eq
 This matrix  $\Pi^{(4)}$ is compatible with two different potential matrices $\Lambda$. The first matrix  is defined by the entry
\[
\Lambda_{12}=q_1^{-a/b+1}\Phi(q_2q_1^{-a/b})\,.
\]
For instance, if $\Phi(z)=z^d$  then
\ben
\Lambda^{(4)}&=&\left(
          \begin{array}{cc}
           -\dfrac{c_1(a+2ad-b)}{2(a+b)}\,q_2^{d+1}\,q_1^{-\frac{a(d+1)}{b}}  & \dfrac{c_1b(d+1)}{a+b}\,q_2^{d}\,q_1^{1-\frac{a(d+1)}{b}} \\
           \\
             -\dfrac{c_1a^2(d+1)}{b(a+b)}\,q_2^{d+2}\,q_1^{-1-\frac{a(d+1)}{b}}&
             \dfrac{c_1(b+2ad+3a)}{2(a+b)})\,q_2^{d+1}\,q_1^{-\frac{a(d+1)}{b}}
          \end{array}
        \right)\,,\nn\\
        \nn\\
        \nn\\
V^{(4)}&=&c_1\,q_2^{d+1}\,q_1^{-a(d+1)/b} \,.   \label{case-4}
\en
It is easy to see that $\mbox{trace}N^2=H_1^2$ and that recursion operator produces  the Hamiltonian only.

In order to get the desired functionally independent integral of motion $H_2$ we have to solve the equation (\ref{def-h2d}). As above, there exist integrals $H_2$  of second, forth or sixth order in momenta. Moreover,  we sometimes  have two independent solutions of the equation (\ref{def-h2d}) associated with different $\varkappa$'s. For instance, integrable system with potentials
 \[
V=q_1^{-2/3}q_2^{-7/3}\,,\nn\\
 \]
has fourth  order polynomial solution of (\ref{def-h2d})  at $\varkappa=1$
\[
H_2^{(4)}=\dfrac{p_1(p_1q_1-p_2q_2)^3}{2}-
\dfrac{13p_1^2q_1^2-44p_1p_2q_1q_2+4p_2^2q_2^2}{16q_1^{2/3}q_2^{4/3}}
+q_1^{-1/3}q_2^{-8/3}
\]
and sixth order polynomial solution at $\varkappa=2$
\ben
H_3^{(4)}&=& 4p_1^2 (p_1q_1 - p_2q_2)^4 + q_1^{-4/3} q_2^{-8/3}\left(10 p_1^2q_1^2 - 16 p_1 p_2 q_1 q_2 + p_2^2 q_2^2\right)\nn\\
&-&4q_1^{-2/3} q_2^{-4/3} p_1(p_1q_1 - p_2q_2) (2p_1^2q_1^2 - 6p_1p_2q_1 q_2 + p_2^2 q_2^2)-\dfrac{3}{q_1 q_2^4}\,.\nn
\en
It can be a first example of a two-dimensional superintegrable system with second, forth and sixth order polynomial integrals of motion, which form non-trivial Poisson algebra \cite{mpt}.

The second potential matrix $\Lambda$, which is compatible with geodesic matrix $\Pi^{(4)}$,   looks like
\ben
\Lambda^{(4')}&=&\left(
          \begin{smallmatrix}
          \frac{c_1(a^2-4ab-b^2)}{2b^2}\,q_2^{\frac{2b}{a-b}}\,q_1^{-\frac{2a}{a-b}} -\frac{2bc_2q_1q_2}{a-b} &c_1\,q_1^{-\frac{a+b}{a-b}}\,q_2^{\frac{3b-a}{a-b}}+c_2q_1^2 \\ \\
           -\frac{a^2c_1}{b^2}\,q_2^{\frac{a+b}{a-b}}\,q_1^{-2-\frac{a+b}{a-b}}  -c_2\,q_2^2&
           \frac{c_1(a^2+4ab-b^2)}{4b^2}\,q_2^{\frac{2b}{a-b}}\,q_1^{-\frac{2a}{a-b}}
           -\frac{2ac_2\,q_1\,q_2}{a-b}
          \end{smallmatrix}
        \right)\,,\nn\\
        \nn\\ \nn\\
V^{(4')}&=&\dfrac{c_1(a^2-b^2)}{2b^2}\,\,q_2^{\frac{2b}{a-b}}\,q_1^{-\frac{2a}{a-b}}
-\dfrac{2c_1(a+b)}{a-b}\,q_1\,q_2\,.
\label{case-4p}
\en
In this case the second integral of motion may be obtained from the recursion operator
\[
H_2^{(4')}=\mbox{trace}\,N^2-H_1^2=
\dfrac{2(a+b)^2}{a-b}(q_1p_1-q_2p_2)^2
-\dfrac{4c_1(a+b)^2}{b^2}\,\,q_1^{-\frac{a+b}{a-b}}\,q_2^{\frac{a+b}{a-b}}\,.
\]

\subsection{Integrals of motion via variables of separation}\label{sep1}

Substituting known separation  coordinates $u=(u_1,\ldots,v_n)$  (\ref{dn-var}) and  momenta $v=(v_1,\ldots,v_n)$  into  separated relations (\ref{seprel}) and solving the resulting equations with respect to $H_1,\ldots,H_n$ one gets a lot of separable Hamiltonians. The main problem is to propose an effective procedure of selection natural Hamiltonians  similar to the Benenti recursion procedure.

The interim problem is to find the  momenta $v=(v_1,\ldots,v_n)$ canonically conjugated to coordinates $u=(u_1,\ldots,v_n)$  (\ref{dn-var}). Different algorithms for explicit computation of  the Darboux-Nijenhuis variables $(u,v)$ are discussed in  \cite{fp02,ts08,ts10,ts10a}.

\vskip0.2truecm
\par\noindent
\textbf{Case 5}: Let us consider geodesic matrix
\bq\label{pi-5}
 \Pi^{(5)}=\left(
      \begin{array}{cc}
       p_2^2 & 0 \\
       \\
        ap_1p_2 &  p_2^2
      \end{array}
    \right)\,,
\eq
which may be endowed with two compatible potential matrices
\[
\Lambda^{(5)}=\left(
                 \begin{array}{cc}
                  c_1q_1^{-\frac{4}{a-2}}  & 0 \\ \\
                  f(q_1)  &  c_1q_1^{-\frac{4}{a-2}}
                 \end{array}
               \right)
\qquad\mbox{and}\qquad
\Lambda^{(5')}=\left(
  \begin{array}{cc}
    c_1q_1^{-\frac{2}{a-2}}+c_2q_1^{-\frac{4}{a-2}} &0  \\ \\
    \dfrac{c_2}{a-2}\,q_1^{-\frac{a+2}{a-2}}\,q_2+f(q_1) &0
  \end{array}
\right)\,.
\]
In  first case, separation coordinates $u_{1,2}$ are the roots of polynomial
\[
B^{(5)}(\lambda)=(\lambda-u_1)(\lambda-u_2)=\lambda^2-2\left(p_2^2+c_1q_1^{-\frac{4}{a-2}}\right)\lambda
+\left(p_2^2-c_1q_1^{-\frac{4}{a-2}}\right)^2\,,
\]
whereas momenta   $v_{1,2}=A(\lambda=u_{1,2})$ are defined by the polynomial
\[
A=\frac{1}{8\left(p_2^2-c_1q_1^{-\frac{4}{a-2}}\right)}\left[\lambda
\left(
\frac{q_2}{p_2}-\frac{(a-2)q_1^{\frac{a+2}{a-2}}p_1}{2c_1}
\right)
-
\frac{q_2(\tau+8p_2^2)}{4p_2}+
(a-2)p_1\left(\frac{\tau q_1^{\frac{a+2}{a-2}}}{8c_1}+q_1\right)
\right]\,,
\]
where $\tau=\mbox{trace}\,N$.   This  first order polynomial  $A(\lambda)$ is the solution of the  auxiliary equations
\bq\label{ab-eq}
\{A(\lambda),B(\mu)\}=\dfrac{B(\lambda)-B(\mu)}{\lambda-\mu}\,,\qquad \{A(\lambda),A(\mu)\}=0\,,
\eq
which ensure that values of $A(\lambda)$ in $u_j$ (\ref{dn-var}) are the desired momenta
\[
v_j=A(u_j)\,,\qquad  \{u_i,v_j\}=\delta_{ij}\,,\quad \{v_i,v_j\}=0\,.
\]
Associated with matrix $\Lambda^{(5')}$ separation
coordinates $u_{1,2}$ are defined by another second order polynomial
\[
B^{(5')}(\lambda)=(\lambda-u_1)(\lambda-u_2)=\lambda^2
-\left(2p_2^2+c_1q_1^{-\frac{2}{a-2}}+c_2q_1^{-\frac{4}{a-2}}\right)\lambda
+p_2^2\left(p_2^2+c_1q_1^{-\frac{2}{a-2}}\right)\,,
\]
for which solution of the equations (\ref{ab-eq}) looks like
\ben
A(\lambda)&=&
\dfrac{1}{2c_1c_2q_1^{-\frac{2}{a-2}}+4c_1p_2^2+c_2^2}
\left[\lambda\left(
\dfrac{q_2\left(2c_2+c_1q_1^{\frac{2}{a-2}}\right)}
{2p_2}
-{(a-2)q_1^{\frac{a+2}{a-2}}p_1}
\right)\right.\nn\\ \nn\\
&+&\left.\dfrac12
\left(
\dfrac{q_2(4c_2+c_1q_1^{\frac{2}{a-2}})\tau+2c_1(c_2q_1^{-\frac{2}{a-2}}+c_1)}{8p_2}
-\dfrac{a-2}{4}(\tau q_1^{\frac{a+2}{a-2}}+2q_1c_2)p_1
\right)
\right]\,.\nn
\en
In  both cases after substituting  $p_{1,2}(u,v)$ into  common  geodesic  Hamiltonian $T$
one gets  separable Hamiltonians, which have the so-called St\"ackel form in $(u,v)$-variables \cite{bl09,fp02,imm00,st}. It allows us to  easily find all the  separable potentials $V(u_1,u_2)$  and to construct additional St\"ackel  integrals of motion. For instance, in  first case, if $a=-1$,   we have
\[
H_1^{(5)}=\dfrac{32c_1^{3/2}}{9}\dfrac{u_1v_1^2+u_2v_2^2}{\sqrt{u_1}-\sqrt{u_2}}
+\dfrac{1}{6}\,(u_1+\sqrt{u_1u_2}+u_2)=\dfrac{p_1^2+p_2^2}{2}
+\dfrac{c_1}{18}\,q_1^{-2/3}\left(3q_1^2+4q_2^2\right)\,.
\]
In  second case, if  $a=-1$ and $c_1=0$, we  have
\[
H_1^{(5')}=\dfrac{8c_2^{3/2}}{9}\dfrac{v_1^2u_1^{3/2}+v_2^2u_2^{3/2}}{u_1-u_2}
+\dfrac{1}{4}(u_1+u_2)=\dfrac{p_1^2+p_2^2}{2}+\dfrac{c_2}{36}\,q_1^{-2/3}(2q_2^2+9q_1^2)\,.
\]
If $a$ is arbitrary,  then we can obtain the same Hamiltonians after some additional canonical transformation of $(p,q)$ variables.

These bi-integrable systems are  the so-called  first and second Holt-like systems  \cite{gdr84}. The third known Holt-like system may be obtained from the Henon-Heiles system (\ref{case-1}).

In generic case the separable geodesic Hamiltonian  $T$ has more complicated non-St\"ackel form and, therefore,  usually we do not know how to get  separable potentials and additional integrals of motion.  Examples of such  generic  Hamiltonians may be found in next Section.

\section{Bi-integrable systems  on  sphere}\label{s-sphere}
\setcounter{equation}{0}
In this Section we consider  geodesic matrices  on a cotangent bundle $T^*\mathbb S^n$ of the sphere $\mathbb S^n$:
\bq\label{p2-sph2}
 P'_T=
 \left(
      \begin{array}{cc}
        \displaystyle \sum_{k=1}^n \mathrm{x}_{jk}(q)\dfrac{\partial \Pi_{jk}}{\partial p_i}-\mathrm{y}_{ik}(q)\dfrac{\partial \Pi_{ik}}{\partial p_j} & \Pi_{ij} \\
        \\
         -\Pi_{ji}\quad&\displaystyle \sum_{k=1}^n\left(\dfrac{\partial \Pi_{ki}}{\partial q_j}-\dfrac{\partial \Pi_{kj}}{\partial q_i}\right)\,\mathrm z_{k}(p\,)\\
      \end{array}
    \right)
\,.
 \eq
 By definition $P'_T$ is the Poisson bivector compatible with canonical ones, so that
\bq\label{w-eq}
[P,P'_T]=[P'_T,P'_T]=0\,.
\eq
It means that the geodesic matrix $\Pi$ and the functions $\mathrm x_{jk}(q),\mathrm y_{ik}(q)$, $\mathrm z_{k}(p)$ are the solutions of these equations. As above, we restrict ourselves by  particular solutions  only.

We present some examples of bi-integrable natural systems on two-dimensional sphere  $\mathbb S^2$, which are related to the rigid body dynamics.  In order to  submit the Hamilton function into the standard form, we will use the vector of angular momentum $J=(J_1,J_2,J_3)$  and  the unit  Poisson vector $x=(x_1,x_2,x_3)$, see \cite{bm05}.

If the   square  integral of motion $p_\psi=(x,J)=0$ is equal to zero,
the rigid body dynamics may be restricted on the sphere  $\mathbb S^2$.
There exists a standard spherical coordinate system  on the cotangent bundle $T^*{\mathbb S}^2$, which consists of Euler angles $\phi,\theta$ and the corresponding momenta $p_\phi,p_\theta$
\[ q=(q_{1},q_{2})=(\phi,\theta)\qquad\mbox{and}\qquad p=(p_1,p_2)=(p_\phi,p_\theta)\,.\]
defined by
\[
\begin{array}{lll}
x_1 =\sin\phi\sin\theta,\qquad& x_2 = \cos\phi\sin\theta,\qquad & x_3 =\cos\theta\\
\\
J_1 =\dfrac{\sin\phi\cos\theta}{\sin\theta}\,p_\phi-\cos\phi\,p_\theta\,,\qquad&
J_2 =\dfrac{\cos\phi\cos\theta}{\sin\theta}\,p_\phi+\sin\phi\,p_\theta\,,
\qquad& J_3 = -p_\phi\,.
\end{array}
\]
We use these variables in definitions of  geodesic Poisson bivectors (\ref{p2-sph2})  and potential parts $\Lambda$
 compatible with them.

\subsection{The Kowalevski top and  Chaplygin system}
  Firstly we consider  a  geodesic bivector $P'_T$ (\ref{p2-sph2}) determined by the degenerate  matrix $\Pi$
\bq\label{kow-pi}
\Pi^{(6)}=\dfrac{1}{\sin^\alpha \theta\, \cos^2\theta}\left(
                                            \begin{array}{cc}
                                              0 & \dfrac{2\,p_\phi\,p_\theta}{\alpha} \\ \\
                                              0 & \cos^2\theta\,p_\phi^2+\sin^2\theta\,p_\theta^2
                                            \end{array}
                                          \right)\,,\qquad \alpha\in \mathbb R\,,
\eq
and by functions
\[\mathrm y_{12}=\cos\theta\Bigl(\sin\theta+\alpha \mathrm x_{22}(\theta)\cos\theta\Bigr)\,,\qquad \mathrm z_{1,2}(p_\phi,p_\theta)=0\,.
\]
Other functions $\mathrm x_{jk}$ and $\mathrm y_{ik}$ are arbitrary. This matrix $\Pi^{(6)}$ is consistent with the following potential matrix
\[
\Lambda^{(6)}=\left(
          \begin{array}{cc}
            a\cos \alpha\phi-b\sin \alpha\phi & \Bigl (a\sin \alpha\phi-b\cos \alpha\phi\Bigr)\cot\theta \\  \\
            \Bigl (a\sin \alpha\phi-b\cos \alpha\phi\Bigr) \tan\theta &  -a\cos \alpha\phi+b\sin \alpha\phi
          \end{array}
        \right)\,,\qquad a,b\in\mathbb R\,.
\]
The eigenvalues of the  corresponding recursion operator $N$ are the  variables of separation, which have been studied in \cite{ts10,ts10a}.  At $\alpha=1,2$  the separable natural Hamiltonians
\bq\label{ham-kc}
H_1=\left(1+\dfrac{1}{\sin^2 \theta}\right)p_\phi^2+p_\theta^2
+2\Bigl(a\cos\alpha \phi-b\sin\alpha\phi \Bigr)\,\sin^\alpha \theta\,,
\eq
have thoroughly familiar forms in physical  variables
\bq\label{h-kow}\begin{array}{l}
H_1^{(6)}=J_1^2+J_2^2+2J_3^2+2ax_2-2bx_1\,,\\ \\
H_1^{(6')}=J_1^2+J_2^2+2J_3^2+2a(x_2^2-x_1^2)-4bx_1x_2\,.
\end{array}
\eq
It is easy to see that they coincide with the Hamilton functions for the Kowalevski top and the Chaplygin system, respectively, see  \cite{bm05} and references within.

The additional integrals of motion are fourth order polynomials in momenta:
 \ben
 H_2^{(6)}&=&-\dfrac{p_\phi^4}{\sin^2\theta}-p_\phi^2p_\theta^2
 -2(a\sin\phi+b\cos\phi)\cos\theta\,p_\theta p_\phi+\sin^2\theta\,(a\sin\phi+b\cos\phi)^2\nn\\
 \nn\\
& &-(a\cos\phi-b\sin\phi)\,\dfrac{2p_\phi^2}{\sin\theta}\,,\nn
 \en
and
\ben
H_2^{(6')}&=&-\dfrac{p_\phi^4}{\sin^2\theta}-p_\phi^2p_\theta^2+\dfrac{2(a\sin 2\phi+b\cos2\phi)\cos^3\theta}{\sin\theta}\,p_\phi\,p_\theta +\sin^4\theta ( a\sin 2\phi+  b\cos 2\phi)^2\nn\\ \nn\\
&-&(a\cos 2\phi-b\sin 2\phi)\left( \dfrac{2-3\cos^2\theta}{\sin^2\theta}\,p_\phi^2+p_\theta^2\right)
+(a^2+b^2)\cos 2 \theta\,.\nn
\en
The corresponding separation relations are non-affine (non-St\"ackel)  relations  \cite{ts10,ts10a}. Note that  these  variables of separation are different from the famous Kowalevski variables.
\begin{rem}
Substituting the same variables of separation into other separation relations we
can obtain different generalizations of bi-integrable Hamiltonians $H_1^{(\alpha)}$ (\ref{h-kow}), see \cite{ts05k,ts10,ts10a}.
\end{rem}

\subsection{The Goryachev-Chaplygin top}
The Goryachev-Chaplygin top  is defined by the Hamilton function
\bq
H_1^{(7)}=J_1^2+J_2^2+4J_3^2+ax_1+\dfrac{b}{x_3^2}=
\dfrac{4-3\cos^2\theta}{\sin^2\theta}\,p_\phi^2+p_\theta^2+c_1\sin\phi\sin\theta+\dfrac{c_2}{\cos^2\theta}\,,
\eq
which is in involution with the  second integral of motion
\[
H_2^{(7)}=2J_3\left(J_1^2+J_2^2+\dfrac{b}{x_3^2}\right)-c_1x_3J_1\,.
\]
This system is a bi-integrable system with respect to  natural bivector $P'$ (\ref{p2-sph2})  defined by
\bq\label{p-gch}
\Pi^{(7)}=\left(
            \begin{array}{cc}
    p_\theta^2 +\dfrac{4-\cos^2\theta}{\sin^2\theta}\,p_\phi^2 & 2p_\phi\,p_\theta \\ \\
             2p_\phi\,p_\theta  &p_\theta^2 -\dfrac{\cos^2\theta}{\sin^2\theta}\,p_\phi^2
            \end{array}
          \right)
\,,\quad
\Lambda^{(7)}=\left(
                \begin{array}{cc}
               \dfrac{c_2}{\cos^2\theta} &0 \\ \\
                  0 &  \dfrac{c_2}{\cos^2\theta}
                \end{array}
              \right)\,,
\eq
 and by the functions
\[
\mathrm x_{22}=\mathrm y_{12}=-\dfrac{\cos\alpha\theta\,\sin\alpha\theta}{\alpha}\,,\qquad\qquad \mathrm z_{k}=\dfrac{p_k}{3}\,.
\]
The corresponding separation variables and some another bi-integrable systems
 separable in these variables are discussed in \cite{ts08k,ts07b}.

\begin{rem}
In \cite{ts09v} we found some linear in momenta Poisson bivectors $\widetilde{P}$ for five  integrable systems on  the sphere $\mathbb S^2$ with cubic additional  integrals of motion. In all these cases quadratic in momenta bivectors $P'=\widetilde{P}P^{-1}\widetilde{P}$  can be rewritten in  natural form (\ref{p2-sph2}) as well.
\end{rem}

Bivectors $\Pi^{(6)} $ (\ref {kow-pi}) and $\Pi^{(7)} $ (\ref{p-gch}) have various $n$-dimensional counterparts on $ T^*\mathbb S^n$ and variant analogs on the cotangent bundles of  other Riemannian manifolds. However, in order to get interesting  integrable systems  with higher order integrals of motion we have to  learn to construct  separable natural  Hamiltonians directly from the variables of separation in  St\"ackel and non-St\"ackel cases.

\section{Generalized  natural Poisson bivectors on $\mathbb R^{2n}$  }\label{sep-gen}
\setcounter{equation}{0}
 In this section we consider some generalizations of natural Poisson  bivector  (\ref{p2-gen}) on $\mathbb R^{2n}$, which are related to various modifications of the geodesic bivector  and  with "potential" parts depending on momenta.

\subsection{Geodesic matrix $\Pi$ depending on coordinates}\label{hh-sep}
 We suppose that $\Pi$ depends on coordinates $q$ and momenta $p$ and
  geodesic bivector $P'_T$ on $\mathbb R^{2n}$ is given by  formulae  (\ref{p2-sph2}). For instance, let us consider  degenerate matrix
\bq\label{pi-8}
 \Pi^{(8)}=\dfrac{1}{2q_1^2}\left(
      \begin{array}{cc}
       2p_1^2 & 0 \\
       \\
        p_1p_2&  0
      \end{array}
    \right)\,,
\eq
which is the similar to matrix $\Pi^{(6)} $ (\ref {kow-pi}) associated with the Kowalevski top and Chaplygin system.
Solving the equations (\ref{w-eq}) with respect to functions $\mathrm x_{j,k}(q), \mathrm y_{i,k}(q)$ and $\mathrm z_{k}(p)$ one gets
\[
\mathrm x_{j1}=\mathrm y_{i1}=-q_1\,,\qquad \mathrm z_{1,2}=0\,.
\]
In generic case potential matrix $\Lambda$ is  compatible with $\Pi^{(6)}$ if and only if
\bq\label{hh-sepL}
\Lambda_{22}=d_1q_2^2+d_2q_2+d_3\,,\qquad d_k\in\mathbb R.
\eq
If $d_1=d_3=0$ and $d_2 = -c_1/4$,  then
\bq\label{l8}
\Lambda^{(8)}=-\dfrac{1}{4}\left(
                \begin{array}{cc}
                  -2c_1q_2-c_2\qquad & -\dfrac{1}{2}c_1q_1-4q_1^{-1}(3c_1q_2^2+2c_2q_2+c_3) \\ \\
                  \dfrac{1}{4} c_1q_1 & c_1q_2
                \end{array}
              \right)\,
\eq
and the  integrals of motion for the Henon-Heiles system $H^{(1)}_1=T+V^{(1)}$ (\ref{case-1})  and $H_2^{(1)}$ (\ref{case-1-h2}) are in involution with respect to the corresponding Poisson bracket.

If $d_1=-c_1$ and $d_2=d_3=0$,  then
\bq\label{l8p}
\Lambda^{(8')}= -\left(\begin{array}{cc}
                 c_3-c_1\left(\dfrac{q_1^2}{2}+2q_2^2\right) \qquad&
                 -c_1q_1q_2-\dfrac{2(2c_1q_2^6-c_3q_2^4+c_2)}{q_1q_2^3}
                  \\ \\
                  -\dfrac{c_1q_1q_2}{2} & c_1q_2^2
                \end{array}
              \right)\,
\eq
and  the integrals of motion for the system with quartic potential  $H^{(1)}_1=T+V^{(1')}$ (\ref{case-1p})  and $H_2^{(1')}$ (\ref{case-1p-h2}) are in involution with respect to the corresponding Poisson bracket.

In  both cases eigenvalues $u_{1,2}$  of   recursion operators $N$  are the variables of separation for the Henon-Heiles system and the system with quartic potential, respectively. The corresponding separated relations are the standard affine  St\"ackel relations, see  \cite{rrg94,ts96}.

\begin{rem}
These variables of separation $u_{1,2}$ have been introduced in \cite{rrg94}  with the help of the singular Painleve' expansions of the solutions of the equations of motion.  Different  properties of these variables of separation and the corresponding  rational Poisson bivectors have been  studied in \cite{ts96, ts06} .

\end{rem}

\subsection{Potential matrix $\Lambda$ depending on momenta}\label{r-toda}
 Open relativistic Toda lattice associated with $\mathcal A_n$ root system is an integrable  system on $\mathbb R^{2n}$ with  following  first two Hamiltonians:
\[
H_1=\sum_{i=1}^n c_i+d_i\,,\qquad
H_2=\sum_{i=1}^{n}\Bigl(\dfrac12(c_i+d_i)^2+c_{i-1}(c_i+d_i)\Bigr)\,,
\]
where
\[
c_i=\exp(q_i-q_{i+1}+p_i)\,,\qquad d_i=\exp(p_i)\,,\qquad
q_0=-\infty,\qquad q_{n+1}=+\infty\,.\]
This system is also known to be bi-Hamiltonian with respect to  second Poisson bracket
 on $\mathbb R^{2n-1}$
  \bq\label{rtoda-pb}
  \{c_k,d_k\}'=c_k\,,\qquad \{ck,d_{k+1}\}'=-c_k\,,\qquad \{d_k,d_{k+1}\}'=c_k\,,
 \eq
  which was found in   \cite{rag89}.  Of course, it is non-natural system in $\mathbb R^{2n}$, but  the corresponding Poisson bivector may be rewritten  in  the generalized natural form, if we  introduce constant matrix
\[
\widehat{E}=
   \alpha\left(
    \begin{array}{cccc}
      1 & 1 & \cdots & 1 \\
      1 &  &  & \vdots \\
      \vdots& \ddots &  & 1 \\
      1 & \cdots &1 & 1
    \end{array}
  \right)- \left(\begin{array}{cccc}
      0 & 0 & \cdots & 0 \\
      1 & 0 &  & \vdots \\
      \vdots&  \ddots &\ddots  & 0 \\
      1 & \cdots &1 & 0
    \end{array}
  \right)\,,\qquad \alpha\in \mathbb R\,,
\]
and antisymmetric matrix $\widehat{A}$ with entries
$\widehat{A}_{i,i+1}=e^{q_i-q_{i+1}-p_{i+1}} $,  so
\[
 \Lambda=-\widehat{E}\,\widehat{A}\,,
\]
similar to the  non-relativistic Toda lattice  (\ref{l-atoda}).  However,  let us emphasize,  that in this case potential matrix $\Lambda$ depends on coordinates and  momenta.  If  we put
\ben
&\Pi=\mbox{diag}\bigl(\exp(-p_1),\ldots,\exp(-p_n)\bigr) \,,\qquad
&( \partial_p \Pi)_{ij} =(\alpha-1)\sum_{k=1}^n \dfrac{\partial \Pi_{ik}}{\partial p_k}
-\alpha\dfrac{\partial\Pi_{jj}}{\partial p_j}\,,
\nn\\
\nn\\
&( \partial_p\Lambda)_{ij} =\displaystyle(\alpha-1)\sum_{k=1}^n \dfrac{\partial\Lambda_{ik}}{\partial p_k}
+\dfrac{\partial \Lambda_{ij}}{\partial p_j}\,,\qquad
&(\partial_q \Lambda)_{ij}=\displaystyle \sum_{k=1}^n\left(\dfrac{\partial \Lambda_{kj}}{\partial q_i}-\dfrac{\partial \Lambda_{ki}}{\partial q_j}\right)\,,
\en
then  $P'$  is a sum of the geodesic Poisson bivector and  potential Poisson bivector
\bq\label{p-todar}
P'= \left(
      \begin{array}{cc}
       \partial_p \Pi &\Pi \\ \\
       -\Pi^\top & 0
      \end{array}
    \right)+ \left(
      \begin{array}{cc}
       \partial_p \Lambda &\Lambda \\ \\
       -\Lambda^\top & \partial_q \Lambda
      \end{array}
    \right)\,.
\eq
As above, it is the Poisson bivector at $\Lambda=0$ and $\Lambda\neq 0$. At any $\alpha$
bivector  (\ref{p-todar}) reduces to the known Poisson bivector (\ref{rtoda-pb}) on $\mathbb R^{2n-1}$  and all the  integrals of motion are the traces of powers of the recursion operator $N=P(P')^{-1}$ \cite{rag89,sur93} .

\begin{rem}
In the similar manner as for non-relativistic Toda lattice we can get natural Poisson bivectors $P'$ for  relativistic Toda lattice associated with other  root systems.
\end{rem}

 \subsection{Additive deformations}\label{hh-gen}
Using canonical transformations and deformations of separated relations we can study more complicated integrable systems.
For instance, let us consider a generalized Henon-Heiles system  \cite{gdr84,h87}   with the potential
 \[
 V=\dfrac{c_1}8\,q_2(3q_1^2+16q_2^2)+c_2\left(2q_2^2
 +\dfrac{q_1^2}{8}\right)+\dfrac{c_4}{q_1^2}+\dfrac{c_5}{q_1^6}\,.
 \]
 and the second integral of motion
 \[
 H_2=H_2^{(1)}+\dfrac{4c_5^2}{q_1^12}
 +\dfrac{c_5(4p_1^2q_1^2+3c_1q_2q_1^4+c_2q_1^4+8c_4)}{q_1^8}
 +\dfrac{c_4(4p_1^2q_1^2+c_1q_2q_1^4+4c_4)}{q_1^4}\,.
 \]
 Here $H_2^{(1)}$ is given by (\ref{case-1-h2}) at $c_3=0$\,.

 Note, that we do not have any additional information  for this system,  as the Lax matrices, $r$-matrices or relations with soliton equations. Nevertheless,  it is easy to directly prove  that these integrals of motion are in bi-involution with respect to the Poisson bracket associated with additive deformation of the natural Poisson bivector $P'$ (\ref{case-1})
 \[
 \widehat{P}=P'+\dfrac{ \sqrt{-2c_5}}{q_1^3}\left(
                  \begin{array}{cccc}
                    0 &  0& p_1-\sqrt{\dfrac{-c_5}{2q_1^6}} &0  \\
                    * & 0 & p_2 & 0 \\
                    *& * & 0 & \dfrac{3c_1}{8}q_1^2+6c_1q_2^2+4c_2q_2 \\
                    * & * & * & 0 \\
                  \end{array}
                \right)\,.
 \]
 This additive deformation may be obtained  from the natural Poisson bivector $P'$ (\ref{case-1}) using trivial canonical transformation
 \bq\label{ctr-hh}
 p_1\to p_1+f(q_1)\,,\qquad \mbox{where}\qquad f(q_1)=-\dfrac{\sqrt{-2c_5}}{q_1^3}\,.
 \eq
  For  generalized  system with quartic potential  \cite{gdr84,h87}
\[
V=\dfrac{c_1}{4}(q_1^4+6q_1^2q_2^2+8q_2^4)+\dfrac{c_2}{2}(q_1^2+4q_2^2)
+\dfrac{2c_3}{q_2^2}+\dfrac{c_4}{q_1^2}+\dfrac{c_5}{q_1^6}\,,
\]
we have to shift known natural bivector $P'$ (\ref{case-1p})  on the similar  additional term
\[
\widetilde{P}=P'+\dfrac{ \sqrt{-2c_5}}{q_1^3}\left(
                  \begin{array}{cccc}
                    0 &  0& p_1-\sqrt{\dfrac{-c_5}{2q_1^6}} & 0 \\
                    * & 0 & p_2 & 0 \\
                    *& * & 0 &8c_1q_2^3+(3c_1q_1^2+4c_2)q_2-\dfrac{4c_3}{q_2^3} \\
                    * & * & * & 0 \\
                  \end{array}
                \right)
 \]
 associated with the same trivial canonical transformation (\ref{ctr-hh}).

We can apply this transformations to the Poisson bivectors $P'$ defined by geodesic matrix $\Pi^{(8)}$ (\ref{pi-8}) and
potential matrices $\Lambda^{(8)}$ (\ref{l8}) and $\Lambda^{(8')} $ (\ref{l8p}) too. In  both cases the "shifted"   variables of separation
 \bq\label{sh-sepvar}
{\tilde u_{1,2}}= u_{1,2}\Bigl(p_1\to p_1+f(q_1)\Bigr)\,,
\eq
 are defined by the initial variables  $u_{1,2}$ obtained in \cite{rrg94} at $c_5=0$.   It is easy to prove that the corresponding separated relations are non-affine  in $H_{1,2}$, i.e. they are non-St\"ackel relations, similar to the Kowalevski top and the generalized Chaplygin system \cite{ts10,ts10a}. These separation relations   will be discussed in the forthcoming publication.

 \begin{rem}
Another example of such additive deformations of the natural Poisson bivectors on the plane and the corresponding separation variables may be found in \cite{mpt}.
 \end{rem}

 \section{Conclusion }
\setcounter{equation}{0}
We address the problem of construction of natural  integrable systems on  Riemannian manifolds $Q$ within the theoretical scheme of bi-Hamiltonian geometry and  introduce the concept of natural Poisson bivectors, which generalizes the Benenti construction of the Poisson bivectors via conformal Killing tensors of gradient type on  $Q$.  We suppose  that  the proposed construction allows us to  describe a  majority of   known  integrable systems with higher order integrals of motion and to find  the corresponding variables of separation in  common framework.

 A lot of known and new integrable systems on  plane  and on  sphere is discussed in detail. These examples may  be useful for  creating a   geometrically  invariant theory, which takes the  constructive answers  to the main open questions:
 \begin{itemize}
  \item how to get  and classify all the natural Poisson bivectors $P'$ on $T^*Q$;
  \item how to describe all the natural Hamilton functions associated with  a given $P'$.
\end{itemize}
 Now we have some particular answers obtained  by direct tedious computations  only.

We hope that the theory of natural Poisson bivectors allows us to investigate   known  $n$-body systems and low-dimensional exotic systems, systems with known Lax representation and systems without it, systems associated with the Killing tensors at $\Pi=0$ and systems with higher order integrals of motion  at $\Pi\neq 0$ and so on.
But,  of course,  this theory can not become a universal panacea and we briefly discuss some possible generalizations and modifications of the natural Poisson bivectors  in  last Section.

The author thanks referees for very useful  suggestions.

\end{document}